\documentclass[10pt, twocolumn]{article}
\usepackage{graphicx} 
\usepackage{amsmath, amssymb, amsthm}
\usepackage{times}
\usepackage{enumitem}
\usepackage{float}
\usepackage[margin=0.75in]{geometry}
\begin{document}
    
    \title{SQUIDs for detection of potential dark matter candidates}
    \author{Siddarth Sivakumar$^1$, Manan Agarwal$^2$, Hannah Rana$^1$\\
    $^1$Center for Astrophysics $|$ Harvard \& Smithsonian, Cambridge, MA, USA\\
   $^2$Anton Pannekoek Institute/GRAPPA, University of Amsterdam, Amsterdam, The Netherlands
    }\date{}

\maketitle
\begin{abstract}
Superconducting QUantum Interference Devices (SQUIDs) are extremely sensitive magnetic flux sensors which render them useful in a wide array of instrumentation. SQUIDs are often paired with other detectors as a readout mechanism to obtain quantitative insight. SQUIDs have impacted many fields but much less addressed is its impact on the field of fundamental physics, particularly in the search for dark matter. Dark matter is believed to make up around $27\%$ of all mass-energy content of the universe and will provide critical insight into understanding large-scale structures of the universe. Axions and WIMPs are the prominent two dark matter candidates whose search has been fuelled by the usage of SQUID readouts. 
\end{abstract}
\section{Introduction}
Superconducting QUantum Interference Devices (SQUIDs), inherently flux-to-voltage transducers, can be used to measure any physical quantity convertible into magnetic flux, used in an array of instrumentation such as signal amplifiers, magnetometers and gradiometers. SQUIDs are considered vital in instrumentation due to their quantum sensitivity (field resolution of 10 \textsuperscript{-17} T \cite{fagaly2006superconducting}) and enormous operational bandwidth. SQUIDs employ Josephson junctions in a superconducting loop; two in direct current (DC) SQUIDs and one in radio frequency (RF) SQUIDs. Due to their dependence on superconductors, they require cryogenic temperatures around 9\,K. Although, recent advancements have brought the advent of high-$T_C$ SQUIDs which have a relatively high operational temperature of 77\,K. SQUID-based applications \cite{doi:https://doi.org/10.1002/3527603646.ch1} extend to various fields; recently even concepts of military applications have been conceived \cite{ouellette1998squid}. SQUIDs are used as magnetometers in biology instrumentation: magnetoencephalography, magnetocardiography and gastro-magnetometry. Geophysical applications include the liquid-nitrogen-cooled SQUIDs that provide insight on low conductivity anomalies at depths from a few metres to tens of kilometres, allowing us to predict the presence and depth of oil and mineral ores. High-T\textbf{\textsubscript{C}} SQUIDs have piqued the interest in Non-Destructive Evaluation (NDE) of materials used to locate flaws that may compromise the integrity of structures later on; it can be used to test aircraft structures, steel-reinforced concrete, and more. Our focus in this review, however, will be their extended usage in particle detection for the detection of dark-matter candidates \cite{clarke2010squids}, an emerging approach in the search for dark matter in fundamental physics. SQUIDs are used as readout amplifiers in the detection of axions and Weakly Interacting Massive Particles (WIMPs), potential dark-matter candidates, as they offer better sensitivity than their traditional counterparts.
\begin{figure}
    \centering
    \includegraphics[width=0.7\linewidth]{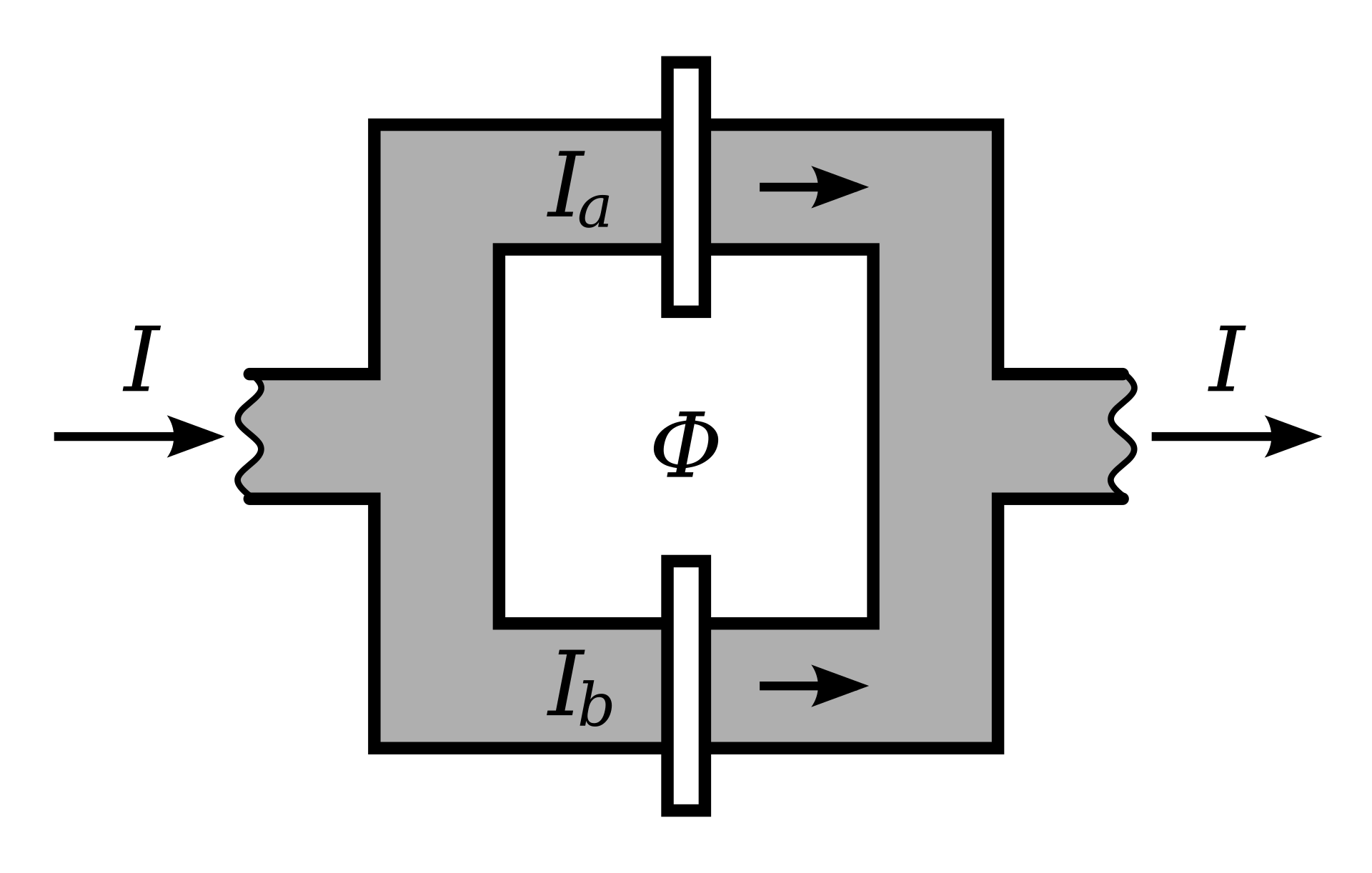}
    \caption{Diagram of the DC SQUID loop \cite{wikipedia_dc_squid_image}}
    \label{fig:1}
    
\end{figure}

\section{Search for Dark Matter}
Dark matter has been theorized to help us explain the formation and evolution of galaxies and large-scale structures in the universe; the existence of dark matter is inferred from various astrophysical observations, such as the rotation curves of galaxies, gravitational lensing, and the cosmic microwave background \cite{clowe2006direct}. It is believed to constitute about 27\% of the universe's mass-energy content. Despite no interaction with light and being detectable only through its gravitational effects, our search for dark matter continues through multiple experiments pushing the boundaries. Researchers have proposed several hypothetical candidates for dark matter particles, with WIMPs and axions being two prominent possibilities.
\begin{figure}[H]
    \centering
    \includegraphics[width=0.5\linewidth]{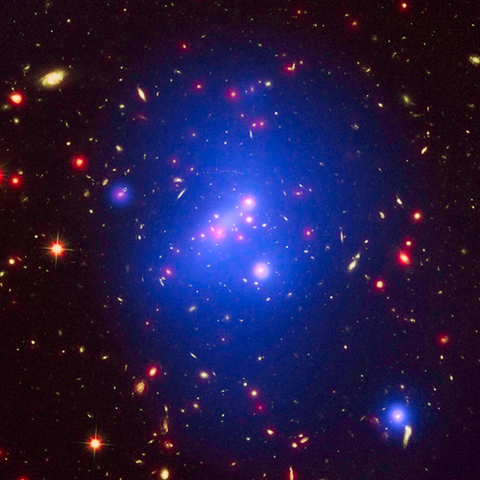}
    \caption{About 90\% of the mass of the galaxy cluster IDCS 1426 is in the form of dark matter. \cite{NIST2016, Brodwin2016}}
    \label{fig:enter-label}
\end{figure}
\begin{enumerate}
    \item \textbf{WIMPs} interact only through the weak nuclear force and gravity, making them difficult to detect. WIMPs are predicted by extensions of the Standard Model of particle physics, such as supersymmetry \cite{bertone2005particle}. Their mass is expected to be in the range of 10 to 1000 times that of a proton (10 to 1000\,GeV/$c^2$), and interact with normal matter at a rate much lower than that of photons or other particles.
    \item \textbf{Axions}, proposed as a solution to the strong CP problem in quantum chromodynamics, interact extremely weakly with normal matter and could contribute to the dark matter density in the universe. They are expected to have masses much lower than WIMPs, potentially in the microelectronvolt ($\mu eV$) range \cite{graham2015experimental}.
\end{enumerate}

\section{ADMX - Axion Detection}
\subsection{Background}
The Axion Dark Matter eXperiment (ADMX) at Lawrence Livermore National Laboratory has been searching for axions since 1996 \cite{asztalos2010squid}. The primary challenge in axion detection is their extremely weak interaction with ordinary matter, requiring highly sensitive detection methods. Traditional detection systems, such as those using the GaAs field-effect transistor (HFET) amplifiers, have inherent noise limitations. The novel approach of using SQUID amplifiers, with their near quantum-limited noise performance, significantly enhances the sensitivity and scan rate of the search. 

\subsection{Experimental Setup}
The experimental setup includes:

\begin{itemize}
    \item A 7.6 Tesla superconducting solenoid with a 0.5-meter diameter bore.
    \item A cylindrical, copper-plated microwave cavity inside the magnet bore.
    \item A DC SQUID amplifier placed above the magnet to minimize interference from the magnetic field.
\end{itemize}
Dark matter axions passing through the cavity are expected to convert into microwave photons, detectable by the resonant cavity. The experiment scanned for axion signals by tuning the cavity frequency and integrating the signal over time to improve the signal-to-noise ratio. The cavity's frequency is tunable by moving copper-plated axial rods, allowing the experiment to scan different axion masses. The expected power from axion-photon conversion, given by the inverse Primakoff effect, is extremely low, requiring ultra-sensitive amplification. The power generated by axion-photon conversion depends on several factors, including the magnetic field strength, cavity volume, axion density, and the quality factors of the cavity and axion signal.
\begin{figure}
    \centering
    \includegraphics[width=1\linewidth]{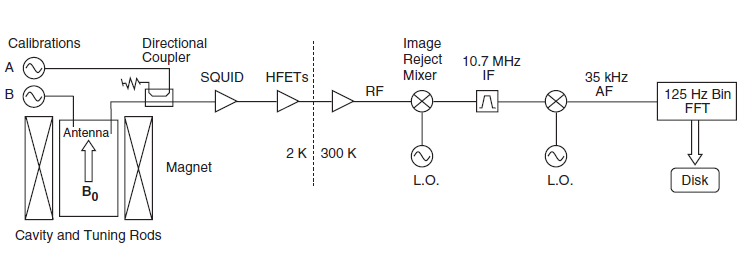}
    \caption{Schematic of the ADMX experiment}
    \label{fig:enter-label}
\end{figure}
\subsection{Role of SQUIDs}

In the ADMX experiment, SQUID amplifiers replace conventional HFET amplifiers, due to several key advantages:

\begin{enumerate}
    \item \textbf{Noise Temperature}: The SQUID amplifier operates with a noise temperature close to the quantum limit, significantly lower than that of HFET amplifiers, allowing for better sensitivity and faster scan rates. The SQUID amplifier has achieved noise temperatures as low as 47\,mK, close to the quantum noise limit of 33\,mK at 700\,MHz.
    \item \textbf{Thermal Stability}: SQUIDs maintain their low noise temperature down to very low physical temperatures, unlike HFET amplifiers which plateau at a few Kelvin.
    \item \textbf{Gain Performance}: The specific design of the SQUIDs used in ADMX includes a resonant microstrip input coil, which mitigates the gain roll-off issue at microwave frequencies.
\end{enumerate}
 \begin{figure}
    \centering
    \includegraphics[width=0.9\linewidth]{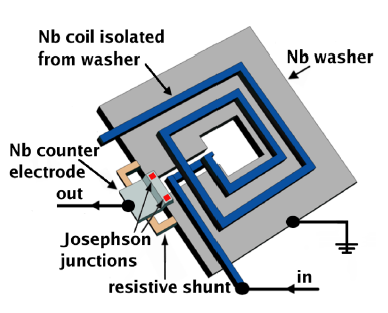}
    \caption{Schematic of a microstrip SQUID amplifier}
    \label{fig:enter-label}
\end{figure}
\subsection{Results and Analysis}
\begin{itemize}[itemsep=0.2em]
    \item \textbf{Noise Reduction}: In the ADMX experiment, the SQUID amplifiers demonstrated a noise temperature of approximately 50\,mK at 1\,GHz, which is close to the quantum limit of 33\,mK. This low noise temperature is crucial for enhancing the signal-to-noise ratio (SNR) and, consequently, the sensitivity of the axion search. 
    \item \textbf{Sensitivity}: The experiment's sensitivity is quantified by the Dicke radiometer equation, which links SNR to system noise temperature, integration time, and bandwidth. With the introduction of SQUID amplifiers, the system noise temperature is minimized, leading to a higher SNR and allowing for a more efficient scan of axion masses.
    \item \textbf{Improvement}: The use of the SQUID amplifier significantly improved the scan rate compared to previous runs using HFET amplifiers.
    \item \textbf{Findings:} The ADMX experiment was able to rule out certain models of dark matter axions with masses between 3.3 and 3.53\,$\mu$eV, with 90\% confidence, assuming a local dark matter density of 0.45 GeV/$cm^3$. This new result extends previous research that had only ruled out axions up to 3.3\,$\mu$eV. While the hunt for dark matter candidates and axions in particular will continue, current findings suggest that they are less likely to have masses in the 3.3 to 3.53\,$\mu$eV range. 
\end{itemize}

\subsection{Future Directions and Conclusion}
Future experiments will use dilution refrigeration to further lower the temperature of the SQUID and cavity, potentially increasing the scan rate and sensitivity. The goal for the future is to cover the entire plausible axion mass range with higher sensitivity, even for the most pessimistic axion-photon couplings. The application of SQUIDs in the ADMX experiment showcases its potential in particle detection instrumentation. The improvements in sensitivity and scan rate afforded by SQUID amplifiers represent a critical step forward in the ongoing quest to detect dark matter axions. 

\section{CRESST - WIMP Detection}

\subsection{Background}
The Cryogenic Rare Event Search with Superconducting Thermometers (CRESST) experiment \cite{angloher2005limits} uses cryogenic detectors that leverage SQUIDs, which play a huge role in enhancing the sensitivity and precision of WIMP detection. The experiment is set up in the Laboratori Nazionali del Gran Sasso (LNGS) underground laboratory in Italy, which provides a shielded environment to minimize background noise from cosmic rays and other sources. 

\subsection{Experimental Setup}
The core idea of the 2004 data run of the CRESST-II phase of the experiment is to identify WIMP interactions by detecting the tiny amounts of energy deposited by the phonon (heat) signal and scintillation light, both of which need to be detected simultaneously and with high precision by sensitive tungsten phase-transition thermometers, when a WIMP collides with a nucleus in the scintillating CaWO4 crystals used as the detector material. The SQUID readout system subsequently measures resistance changes that allow us to infer changes in temperature. The setup includes:
\begin{figure}
    \centering
    \includegraphics[width=0.75\linewidth]{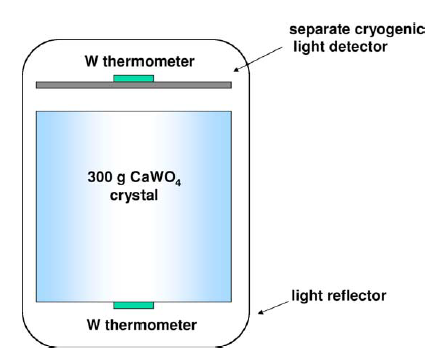}
    \caption{Schematic of the detector module}
    \label{fig:enter-label}
\end{figure}
\begin{itemize}
    \item \textbf{Detectors}: Two 300\,g CaWO4 scintillating crystals were used as targets in their respective modules with reflecting housings, with associated cryogenic light detectors that use the phonon-light technique, which measures both phonon and light signals to identify nuclear recoils indicative of WIMP interactions. Using both channels help distinguish nuclear recoils from background events.
\end{itemize}

\begin{itemize}
    \item \textbf{Phonon Channel}: The phonon signal is the energy transfer to a nucleus from a WIMP-nucleus elastic scattering event, which is measured using a cryogenic calorimeter read out by a SQUID. The amount of energy deposited by a WIMP is expected to be very low and SQUID's sensitivity allows it to detect the minute temperature changes resulting from the phonon signal with high accuracy and precision.
    \item \textbf{Light Channel}: Alongside the phonon channel, a cryogenic light detector measures the scintillation light. This dual measurement system—phonon and light—discriminates between nuclear recoils (potential WIMP signals) and electron recoils (background noise) and enables the identification and suppression of background events, improving the signal-to-noise ratio crucial for WIMP detection. 

\end{itemize}

\begin{itemize}
    \item \textbf{Readout}: The system uses SQUID readout electronics to measure resistance changes in the tungsten thermometers attached to the detectors.
    \item \textbf{Stability}: Temperature regulation and periodic calibration pulses ensure stability and accurate energy measurements throughout the experiment.

\subsection{Role of SQUIDs}
    \item \textbf{Sensitivity}: SQUIDs provide unmatched sensitivity in detecting the minute signals produced by WIMP interactions.
    \item \textbf{Noise Reduction}: The high precision of SQUIDs helps in significantly reducing background noise, which is crucial for identifying rare events.
    \item \textbf{Dual-channel Detection}: By enabling simultaneous detection of phonons and scintillation light, SQUIDs enhance the ability to discriminate between different types of interactions, improving overall detection accuracy.

\end{itemize}

\begin{figure}
    \centering
    \includegraphics[width=1\linewidth]{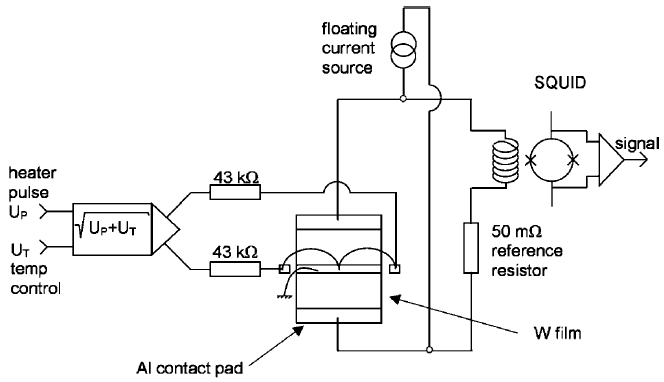}
    \caption{Schematic of the phonon channel readout}
    \label{fig:enter-label}
\end{figure}
\subsection{Results and Analysis}

\begin{itemize}
    \item \textbf{Data Collection}: Data was collected from January 31 to March 23, 2004, with a net exposure of 20.5\,kg days.
    \item \textbf{Nuclear Recoil Events}: A total of 16 events in the nuclear recoil acceptance region (12-40\,keV) were observed, corresponding to a rate of (0.87 $\pm$ 0.22) events/(kg day), which is consistent with expected neutron background. This demonstrates the effectiveness of the SQUID-based detection system in distinguishing between neutron-induced recoils and potential WIMP signals. 
    \item \textbf{Background Suppression}: The phonon-light technique provides strong suppression of non-nuclear recoil backgrounds, enhancing the experiment's sensitivity to WIMPs. The combination of phonon and light signals from the scintillating cryogenic detector allowed for a strong suppression of non-nuclear recoil backgrounds, enabling accurate identification of potential WIMP interactions.
    \item \textbf{Quenching Factor:} It is the measure of the reduction in light yield for nuclear recoils compared to electron/gamma events. For CaWO4, the quenching factor was found to be 7.4 for neutrons and 40 for tungsten recoils.
    \item \textbf{Findings}: The experiment set significant limits on the WIMP-nucleon cross-section. The experiment set exclusion limits on the same, with the best module showing no events in the region corresponding to tungsten recoils. Hence, no events could be conclusively attributed to WIMPs and the search for dark matter candidates continues.
\end{itemize}

\subsection{Future Directions and Conclusion}
Challenges remain, such as the need for extremely low temperatures and the complexity of the SQUID system. Future upgrades will include additional shielding and more detector modules to further reduce background and improve sensitivity, with a focus on enhancing the resolution of the light detectors and studying quenching factors to better understand WIMP interactions. The application of SQUIDs in the CRESST-II phase of the experiment showcases its potential in particle detection instrumentation. The improvements in sensitivity and background suppression due to SQUIDs represent a critical step forward in the ongoing quest to detect WIMPs.



\section{Conclusion}
The integration of SQUIDs has significantly advanced dark matter research by enhancing the detection of weakly interacting massive particles (WIMPs) and axions. In the ADMX experiment, SQUID amplifiers achieved near quantum-limited noise temperatures, improving sensitivity and scan rates, and ruling out axion models with masses between 3.3 and 3.53\,$\mu$eV. In the CRESST experiment, SQUID readouts enabled precise dual-channel detection of phonon and scintillation light signals, improving background suppression and setting exclusion limits on the WIMP-nucleon cross-section. \textbf{These advancements underscore SQUIDs' critical role in reducing noise and increasing detection accuracy, showing great promise for facilitating future discoveries in the search for dark matter and bringing us closer to understanding the universe's unseen mass.}
\section{Acknowledgements}
Special thanks goes to our supervisors H. Rana and M. Agarwal for their incredible mentorship and moral support, and to the CCIR team for giving me an opportunity to connect with incredible people and provide extensive support.

 \bibliographystyle{unsrt}   
 \bibliography{mybibz}

\begin{thebibliography}{10}

\bibitem{fagaly2006superconducting}
RL~Fagaly.
\newblock Superconducting quantum interference device instruments and applications.
\newblock {\em Review of scientific instruments}, 77(10), 2006.

\bibitem{doi:https://doi.org/10.1002/3527603646.ch1}
Alex~I. Braginski and John Clarke.
\newblock {\em Introduction}, chapter~1, pages 1--28.
\newblock John Wiley \& Sons, Ltd, 2004.

\bibitem{ouellette1998squid}
Jennifer Ouellette.
\newblock Squid sensors penetrate new markets.
\newblock {\em Ind. Phys.}, 4(2):20--23, 1998.

\bibitem{clarke2010squids}
John Clarke.
\newblock Squids: Then and now.
\newblock {\em International journal of modern physics B}, 24(20n21):3999--4038, 2010.

\bibitem{wikipedia_dc_squid_image}
{Wikipedia contributors}.
\newblock {DC SQUID Diagram}, 2023.
\newblock [Accessed: 17-Jul-2024].

\bibitem{clowe2006direct}
Douglas Clowe, Maru{\v{s}}a Brada{\v{c}}, Anthony~H Gonzalez, Maxim Markevitch, Scott~W Randall, Christine Jones, and Dennis Zaritsky.
\newblock A direct empirical proof of the existence of dark matter.
\newblock {\em The Astrophysical Journal}, 648(2):L109, 2006.

\bibitem{NIST2016}
NIST.
\newblock Shedding light on dark matter with squids.
\newblock {\em NIST News}, 2016.
\newblock Accessed: 2024-07-18.

\bibitem{Brodwin2016}
Mark Brodwin, Michael McDonald, Anthony~H. Gonzalez, S.~A. Stanford, Peter~R. Eisenhardt, Daniel Stern, and Gregory~R. Zeimann.
\newblock Idcs j1426.5+3508: The most massive galaxy cluster at z $>$ 1.5.
\newblock {\em The Astrophysical Journal}, 817(2):122, 2016.

\bibitem{bertone2005particle}
Gianfranco Bertone, Dan Hooper, and Joseph Silk.
\newblock Particle dark matter: Evidence, candidates and constraints.
\newblock {\em Physics reports}, 405(5-6):279--390, 2005.

\bibitem{graham2015experimental}
Peter~W Graham, Igor~G Irastorza, Steven~K Lamoreaux, Axel Lindner, and Karl~A van Bibber.
\newblock Experimental searches for the axion and axion-like particles.
\newblock {\em Annual Review of Nuclear and Particle Science}, 65(1):485--514, 2015.

\bibitem{asztalos2010squid}
Stephen~J Asztalos, G~Carosi, C~Hagmann, D~Kinion, K~Van~Bibber, M~Hotz, LJ~Rosenberg, G~Rybka, J~Hoskins, J~Hwang, et~al.
\newblock Squid-based microwave cavity search for dark-matter axions.
\newblock {\em Physical review letters}, 104(4):041301, 2010.

\bibitem{angloher2005limits}
G~Angloher, C~Bucci, P~Christ, C~Cozzini, F~Von~Feilitzsch, D~Hauff, S~Henry, Th~Jagemann, J~Jochum, H~Kraus, et~al.
\newblock Limits on wimp dark matter using scintillating cawo4 cryogenic detectors with active background suppression.
\newblock {\em Astroparticle Physics}, 23(3):325--339, 2005.

\end{thebibliography}

\end{document}